\documentclass[sigconf]{acmart}

\usepackage[subtle]{savetrees}

\usepackage{acronym}
\usepackage{bbm}
\usepackage{booktabs}
\usepackage[skip=0pt]{caption}
\usepackage{subcaption}
\usepackage{paralist}
\usepackage{pgfplots}
\usepackage[group-separator={,}]{siunitx}
\usepackage{tikz} 
\usepackage[normalem]{ulem}
\usepackage{balance}

\usepackage{enumitem}

\newcommand{\nonsequentialsymbol}{\downarrow\mathrel{\mspace{-1mu}}\uparrow}

\acrodef{AUC}{\emph{area under curve}}
\acrodef{bi-RNN}{\emph{bidirectional recurrent neural network}}
\acrodef{DBN}{\emph{dynamic Bayesian network}}
\acrodef{DCM}{\emph{dependent click model}}
\acrodef{CCM}{\emph{click-chain model}}
\acrodef{CSM}{click sequence model}
\acrodef{ECDF}{\emph{empirical cumulative distribution function}}
\acrodef{EOS}{\emph{end of sequence}}
\acrodef{GRU}{\emph{gated recurrent unit}}
\acrodef{NCM}{\emph{neural click model}}
\acrodef{PGM}{\emph{probabilistic graphical model}}
\acrodef{RNN}{recurrent neural network}
\acrodef{SERP}{search engine result page}
\acrodef{SGD}{\emph{stochastic gradient descent}}
\acrodef{UBM}{\emph{user browsing model}}

\title{A Click Sequence Model for Web Search}

\author{Alexey Borisov}
\orcid{}
\affiliation{\institution{Yandex \& University of Amsterdam}\city{Moscow}\country{Russia}}
\email{alborisov@yandex-team.ru}

\author{Martijn Wardenaar}
\orcid{}
\affiliation{\institution{University of Amsterdam}\city{Amsterdam}\country{The Netherlands}}
\email{martijnwardenaar@gmail.com}

\author{Ilya Markov}
\orcid{}
\affiliation{\institution{University of Amsterdam}\city{Amsterdam}\country{The Netherlands}}
\email{i.markov@uva.nl}

\author{Maarten de Rijke}
\orcid{0000-0002-1086-0202}
\affiliation{\institution{University of Amsterdam}\city{Amsterdam}\country{The Netherlands}}
\email{derijke@uva.nl}



\copyrightyear{2018}
\acmYear{2018}
\setcopyright{acmlicensed}
\acmConference[SIGIR'18]{The 41st International ACM SIGIR Conference on Research \& Development in Information Retrieval}{July 8--12, 2018}{Ann Arbor, MI, USA}
\acmBooktitle{SIGIR'18: The 41st International ACM SIGIR Conference on Research \& Development in Information Retrieval, July 8--12, 2018, Ann Arbor, MI, USA}
\acmPrice{15.00}
\acmDOI{10.1145/3209978.3210004}
\acmISBN{978-1-4503-5657-2/18/07}

\begin{abstract}

Getting a better understanding of user behavior is important for advancing information retrieval systems.
Existing work focuses on modeling and predicting single interaction events, such as clicks.
In this paper, we for the first time focus on modeling and predicting sequences of interaction events. And in particular, sequences of clicks.

We formulate the problem of click sequence prediction
and propose a \acf{CSM} that aims to predict the order in which a user will interact with search engine results.
\ac{CSM} is based on a neural network that follows the encoder-decoder architecture.
The encoder computes contextual embeddings of the results.
The decoder predicts the sequence of positions of the clicked results.
It uses an attention mechanism to extract necessary information about the results at each timestep.
We optimize the parameters of \ac{CSM} by maximizing the likelihood of observed click sequences.

We test the effectiveness of \ac{CSM} on three new tasks:
\begin{inparaenum}[(i)]
    \item predicting click sequences,
    \item predicting the number of clicks, and
    \item predicting whether or not a user will interact with the results in the order these results are presented on a \ac{SERP}.
\end{inparaenum}
Also, we show that \ac{CSM} achieves state-of-the-art results on a standard click prediction task, where the goal is to predict an unordered set of results a user will click on.

\end{abstract}

\begin{document}
\maketitle


\section{Introduction}\label{sec:introduction}
Search engines play an important role in our everyday lives.
One way to improve them is by getting a better understanding of user search behavior, such as clicks, dwell times, mouse movements, etc.
So far, models of user behavior have focused on modeling and predicting single events, e.g., clicks~\citep{chuklin2015click} and mouse movements~\citep{huang-clicks-2011}, and properties of these events, e.g., time between clicks~\citep{borisov2016context}.
In this paper for the first time we focus on modeling and predicting sequences of information interaction events
and, in particular, sequences of clicks.

Although people tend to make only one (or sometimes no) click on a \acf{SERP},
multi-click query sessions constitute a significant part of search traffic.
For example, about 23\% of the query sessions in the Yandex relevance prediction challenge dataset
contain multiple clicks (see \S\ref{sec:data} for details).
It is commonly assumed that users traverse search results from top to bottom,
which leads to the assumption that clicks are ordered by the position of search results.
However, it was shown that in practice this assumption does not always hold,
and that up to 27.9\%--30.4\% of multi-click sequences, depending on the dataset, are not ordered by position~\cite{wang2015incorporating}.

We aim to create tools that help us understand, model and predict sequences of clicks on search engine results, which is important because it provides an opportunity for improving the user search experience.
For example, knowing that a user is likely to click on many results or that there are high chances that the user will interact with the results in an order other than the one in which the results are presented on a \ac{SERP} can be used by a search engine to proactively show an advice or make a change in the ranking.

We propose a \acfi{CSM} that predicts a probability distribution over click sequences.
At the core of our model is a neural network with encoder-decoder architecture.
We implement the encoder as a \ac{bi-RNN} that goes over the search engine result page from top to bottom and from bottom to top and outputs contextual embeddings of the results.
We implement the decoder as a \ac{RNN} with an attention mechanism.
The decoder is initialized with the final states of the forward and backward \acp{RNN} of the encoder.
It is used to predict the sequence of positions of the clicked results.
The whole network is trained by maximizing the likelihood of the observed click sequences.

We evaluate our proposed \ac{CSM} using a publicly available click log and show that
\ac{CSM} provides good means to generate a short list of $K$ click sequences that contains the observed click sequence with a high probability.
We present an analysis of the performance of \ac{CSM} for query sessions with different numbers of clicks and query sessions in which clicks are ordered/not ordered by position.
We measure the performance of \ac{CSM} on two new tasks: predicting the number of clicks and predicting ordered/unordered sequences of clicks.
Additionally, we show that \ac{CSM} achieves state-of-the-art results on the standard click prediction task, which allows us to compare \ac{CSM} to traditional click models that model and predict single events, namely clicks.

Overall, we make the following contributions:
\begin{itemize}
    \item We formulate a novel problem of predicting click sequences.
    \item To solve this problem, we propose a \acf{CSM} based on neural networks.
    \item We evaluate \ac{CSM} on a range of prediction tasks,
        namely predicting click sequences, predicting the number of clicks, predicting ordered/unordered sequences of clicks
        and, finally, predicting clicks themselves.
\end{itemize}

\noindent%
As to the potential impact of the proposed \ac{CSM} model, we believe it can be used to predict that
\begin{inparaenum}[(i)]
    \item a user will click on more than one result, which may indicate that a user has a complex information need~\citep{kravi-one-2016}; or that
    \item a user will interact with the results not in the order in which these results are presented on the SERP, which may indicate that a user is struggling and there are problems in the ranking of the results~\citep{odijk-struggling-2015}.
\end{inparaenum}
\ac{CSM} can help us identify queries for which there is a room for improvement (in terms of user experience) and it can serve as a quick analysis tool to interpret how a particular change in the ranking of the results will influence user click behavior.

The rest of the paper is structured as follows. 
In Section~\ref{sec:problem statement} we provide a precise statement of the click sequence modeling and prediction problems that we are tackling.
Section~\ref{sec:method} introduces our neural network based model for predicting click sequences.
In Section~\ref{sec:experimental setup} we describe the setup of our experiments and Section~\ref{sec:results} presents the results of those experiments.
We describe related work in Section~\ref{sec:related work} and conclude in Section~\ref{sec:conclusion and future work}.

\section{Problem Statement}\label{sec:problem statement}

In this section, we formulate the problem of click sequence prediction (\S\ref{sec:problem}) and propose three prediction tasks that can be solved by a model capable of predicting click sequences (\S\ref{sec:prediction tasks}).

\subsection{Problem}\label{sec:problem}
Since the number and order of clicks may vary even for the same query and ranking of results (e.g., due to different users and contexts), there exists no unique \emph{correct} click sequence, but a (possibly infinite) set of \emph{probably correct} sequences does exist.
Therefore, the main goal of this paper is to build a model that, given a query and a ranking of results,
describes these probably correct click sequences.

To achieve this goal, we define the \emph{click sequence prediction problem} as follows.
First, we learn a probability distribution $\mathcal{M}$ over all possible click sequences.
Second, we use this learned distribution to obtain the $K$ most probable click sequences.
These $K$ sequences are then used to reason about the properties of the set of probably correct sequences mentioned above, e.g., predicting the expected number of clicks, the expected order of clicks, etc.

More formally, we define a click sequence model $\mathcal{M}$ as follows:
\begin{equation}
   \mathcal{M}: \quad P(s \mid q, r_1, \dots, r_N),
   \label{eq:problem}
\end{equation}
where $q$ is a query, $r_1, \dots, r_N$ is an ordered list of results and $s$ is a sequence of positions of the clicked results $(p_1, \dots, p_{S})$.

\subsection{Prediction tasks}\label{sec:prediction tasks}
There are many possible applications for a model~$\mathcal{M}$ that is capable of
\begin{inparaenum}[(i)]
    \item predicting a probability distribution over click sequences (Eq.~\ref{eq:problem}), and
    \item retrieving the $K$ most probable click sequences.
\end{inparaenum}
It can be used to simulate user behavior, which is important in \emph{online learning to rank} research~\citep{hofmann2013balancing,schuth2016multileave},
or as a tool for analyzing how a particular change in the ranking of results will influence user click behavior.
However, we do not investigate these applications in this work.
Instead, we address three tasks that are both practically useful and help to evaluate the performance of the model~$\mathcal{M}$.

\smallskip\noindent
\textbf{Task 1 (predicting the number of clicks).}
The goal of this task is to predict on how many results a user will click.
Clicking on more than one result might indicate that a user has a complex information need.
Clicking on more than three or four results might indicate that a user is struggling or doing an exhaustive search~\citep{hassan-awadallah-struggling-2014,odijk-struggling-2015}.
Both signals can be used by a search system to proactively show an advice or make a change in the ranking.
Thus, we formally define Task~1 as predicting whether a user will click on $\le L$ results.

To estimate the probability of clicking on $\le L$ results,
we generate a large number $K$ (e.g., $K = 1024$) most probable click sequences $s_1, \dots, s_{K}$ and marginalize over those sequences that have $\le L$ clicks:
\begin{equation}
    P(|s| \le L) = \sum_{s \in \{s_1, \dots, s_K\}} P(s) \mathbbm{1}[|s| \le L].
    \label{eq:estimating probability of <= L clicks}
\end{equation}

\smallskip\noindent
\textbf{Task 2 (predicting non-consecutive click sequences).}
The goal of this task is to predict whether a user will interact with results in the order these results are presented on a \ac{SERP} or in a different order, which we refer to as a \textit{non-consecutive} order.
Interacting with results in a non-consecutive order might indicate that a user is struggling~\citep{scaria-last-2014}.
As mentioned in Task~1, such a signal can be used by a search engine to proactively show an advice or make a change in the ranking.

Similarly to Task 1, we estimate the probability of clicking on results in a non-consecutive order by summing probabilities of the $K$ most probable click sequences $s_1, \dots, s_K$ according to $\mathcal{M}$ in which a user clicks on a result $r_i$ after clicking on a result $r_j$ located below $r_i$ ($i < j$):
\begin{equation}
    P(\nonsequentialsymbol) = \sum_{s \in \{s_1, \dots, s_K\}} P(s) \mathbbm{1}[s \text{ is non-consecutive}].
    \label{eq:estimating probability that click sequence will be ordered by position}
\end{equation}

\smallskip\noindent
\textbf{Task 3 (predicting clicks).}
The last task is actually a standard task solved by click models~\citep{chuklin2015click}.
The goal is to predict a subset of the presented results~$r_1, \dots, r_{N}$ on which a user will click.
Being able to predict that a user will not interact with a subset of results, opens the door for reranking~\citep{yandex-personalized-2014}.

Similarly to Task 1 and Task 2, we estimate the click probability for position $p$ by summing probabilities of the $K$ most probable click sequences $s_1, \dots, s_K$ according to $\mathcal{M}$ in which a user clicks on that position:
\begin{equation}
    P_{\text{click}}(p \mid q, r_1, \dots, r_N) = \sum_{s \in \{s_1, \dots, s_K\}} P(s) \mathbbm{1}[p \in s].
    \label{eq:estimating unconditional click probabilities}
\end{equation}
In practice, it is probably better to use simpler models to predict clicks.
So we use this task mainly to compare with existing work.
We expect the results for a good model~$\mathcal{M}$ to be not much worse compared to the results for click models specifically developed  for this task~\cite{chuklin2015click}.
In fact, as we show in \S\ref{sec:results}, \ac{CSM} achieves state-of-the-art performance on this task.


\section{Method}\label{sec:method}
In this section we propose the \acfi{CSM}, a model for predicting click sequences.
We use $s$ to denote a sequence of positions of the clicked results with a special \ac{EOS} token appended to it, i.e., $s = (p_1, \dots, p_k, \text{EOS})$.

\ac{CSM} is a neural network that is trained to maximize the likelihood of observed click sequences:
\begin{equation}
    \mathcal{L}(s_1, \dots, s_{|S|}) \rightarrow \max_{\Theta},
    \label{eq:maximize_likelihood}
\end{equation}
where $\Theta$ denotes the network parameters and $S = (s_1, \dots, s_{|S|})$ denotes click sequences used for training.

\begin{figure}[h!]
    \includegraphics[width=\linewidth]{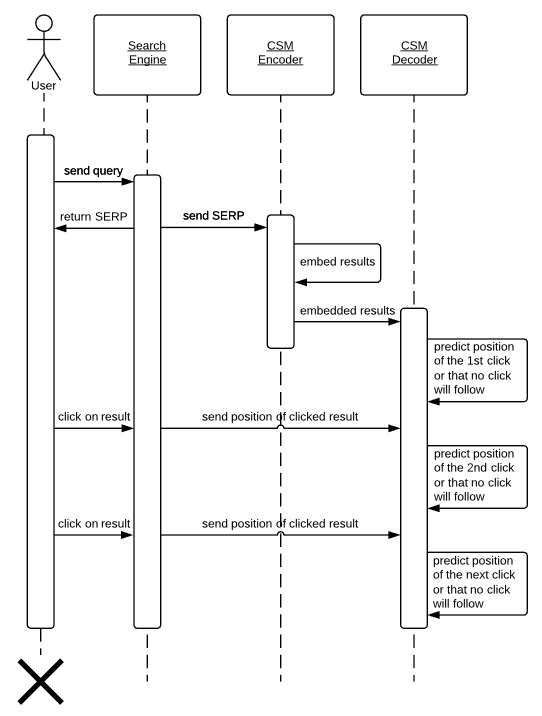}
    \caption{Modeling click sequences with \ac{CSM}.}
    \label{fig:sequence_diagram}
\end{figure}

The network consists of two parts, called \emph{encoder} and \emph{decoder}.
The encoder takes a user's query~$q$ and a list of search engine results $r_1, \dots, r_N$ as input and computes embeddings of the results, $\mathbf{r}_1, \dots, \mathbf{r}_N$.
The embedded results $\mathbf{r}_1, \dots, \mathbf{r}_N$ are passed to the decoder,
which at each timestep $t=0, 1, \dots$ outputs a probability distribution over $N+1$ positions.
Positions $1, \dots, N$ correspond to clicking on the results $r_1, \dots, r_N$.
The $(N + 1)$-th position corresponds to predicting that there will be no clicks (\ac{EOS}).
Upon observing a click at timestep~$t$, the decoder updates its current state using the position $p_t$ of the clicked result.
Figure~\ref{fig:sequence_diagram} illustrates the workflow in the form of a UML sequence diagram.

In \S\ref{sec:network_architecture} we discuss the implementation of the encoder and decoder.
Then, in \S\ref{sec:beam_search} we explain how to achieve the main goal of this study, i.e., predict $K$ most probable click sequences using \ac{CSM}.
Finally, in \S\ref{sec:training} we specify training details.

\subsection{Network architecture}\label{sec:network_architecture}
\smallskip\noindent
\textbf{Encoder.}
The aim of the encoder is to obtain information from a user's query~$q$ and results~$r_1, \dots, r_N$ presented on a \ac{SERP}, and pass this information to the decoder.
We represent the encoded information as a list of embeddings~$\mathbf{q}, \mathbf{r}_1, \dots, \mathbf{r}_N$, where each result embedding~$\mathbf{r}_i$ should contain:%
\begin{inparaenum}[(i)]
    \item information about the result~$r_i$,
    \item the results surrounding $r_i$ and
    \item the query~$q$.
\end{inparaenum}
Below we sometimes use $\mathbf{r}_0$ instead of $\mathbf{q}$ to simplify the notation.

We propose to implement the encoder as a bidirectional recurrent neural network,
which goes over the \ac{SERP} in the top down order, i.e., $q, r_1, \dots, r_N$, and in the reverse order, i.e., $r_N, \dots, r_1, q$.
The first, \emph{forward RNN}, produces embeddings $\overrightarrow{\mathbf{q}} (=\overrightarrow{\mathbf{r}_{0:0}}), \overrightarrow{\mathbf{r}_{0:1}}, \dots, \overrightarrow{\mathbf{r}_{0:N}}$.
The second, \emph{backward RNN}, produces embeddings $\overleftarrow{\mathbf{r}_{N:N}}$, \dots, $\overleftarrow{\mathbf{r}_{N:1}}$, $\overleftarrow{\mathbf{q}} (= \overleftarrow{\mathbf{r}_{N:0}})$.
These embeddings are concatenated to form the final embeddings $\mathbf{q} (=\mathbf{r}_0), \mathbf{r}_1, \dots, \mathbf{r}_N$ produced by the encoder.

We represent $q, r_1, \dots, r_N$ using the best performing behavioral features proposed in~\citep{borisov2016neural}.
These features count the number of times a particular click pattern, i.e., a set of positions of the clicked results, was observed on a \ac{SERP}.
A query $q$ is represented as a $2 ^ N$ dimensional vector, where each component counts the number of times a click pattern was observed in query sessions generated by $q$.
The representation of a search result $r$ consists of two parts, both of size $N 2 ^ N$.
The components of the first part count, for each position $p=1, \dots, N$, the number of times a click pattern was observed in query sessions in which $r$ appears on position $p$.
The components of the second part are similar, but include only query sessions generated by $q$.
We apply a linear transformation to these sparse behavioral features to obtain the embeddings $\mathbf{x}_0, \dots, \mathbf{x}_N$ of $q, r_1, \dots, r_N$, which are passed to the \acp{RNN}.

We describe the encoder formally using Eqs.~\ref{eq:encoder_features}--\ref{eq:encoder_concatenation}:
\begin{eqnarray}
    \mathbf{x}_i &=& \begin{cases} \text{Embed}(q) & i=0 \\ \text{Embed}(r_i) & i=1, \dots, N \end{cases}
    \label{eq:encoder_features}
    \\
    \overrightarrow{\mathbf{r}_{0:0}}, \dots, \overrightarrow{\mathbf{r}_{0:N}} &=& \text{RNN}_{\text{forward}}(\mathbf{x}_0, \dots, \mathbf{x}_N) \label{eq:encoder_forward_rnn}
    \\
    \overleftarrow{\mathbf{r}_{N:N}}, \dots, \overleftarrow{\mathbf{r}_{N:0}} &=& \text{RNN}_{\text{backward}}(\mathbf{x}_0, \dots, \mathbf{x}_N) \label{eq:encoder_backward_rnn}
    \\
    \mathbf{r}_i &=& [\overrightarrow{\mathbf{r}_{0:i}}, \overleftarrow{\mathbf{r}_{N:i}}] \qquad (i = 0, \dots, N) \label{eq:encoder_concatenation}
\end{eqnarray}

\smallskip\noindent
\textbf{Decoder.}
The aim of the decoder is to predict a probability distribution over $(N+1)$ positions at each timestep.
(As mentioned at the start of \S\ref{sec:method}, the $(N+1)$-th position corresponds to predicting that there will be no clicks.)
To make a good prediction at timestep $(t+1)$, we need to incorporate into the decoder the information about the position $p_t$ of the result clicked at timestep~$t$.

We propose to implement the decoder as an \ac{RNN} that at each timestep~$t=0, 1, \dots$ outputs a vector~$\mathbf{o}_t$ used to predict the probability distribution, and updates its hidden state using the position~$p_t$ of the observed click at timestep~$t$.
We also use an attention mechanism~\citep{bahdanau2014neural} to help the decoder extract the most relevant information from the list of embeddings~$\mathbf{r}_0, \dots, \mathbf{r}_N$ at each timestep.

We initialize the hidden state of the decoder \ac{RNN} using the concatenation of the final states of the forward and backward \ac{RNN}s of the encoder, $[\overrightarrow{\mathbf{r}_{0:N}}, \overleftarrow{\mathbf{r}_{N:0}}]$, passed through a linear transformation; we use $W_\text{init}$ to denote the transformation matrix.
To obtain the probability distribution over $(N+1)$ positions at timestep~$t$, we concatenate the vector $\mathbf{o}_t$ predicted by the decoder \ac{RNN} and the attention vector $\mathbf{a}_t$ computed at timestep~$t$, and pass the result through a linear transformation $W_\text{output}$ followed by softmax.\footnote{
Using the output of an \ac{RNN} together with an attention vector has been shown to improve prediction performance~\citep{bahdanau2014neural, pascanu2013construct}.
In the literature, this idea is known as \emph{deep output}~\citep{pascanu2013construct}.}
We represent the position~$p_t$ of the observed click at timestep~$t$ as a one-hot vector~$\mathbf{p}_t$ of size~$N$.
And apply a linear transformation~$W_{pos}$ to it before passing to the decoder \ac{RNN}.

We describe the decoder formally using Eqs.~\ref{eq:decoder_init}--\ref{eq:decoder_softmax}.
\begin{eqnarray}
    \mathbf{s}_0 &=& W_{\text{init}} [\overrightarrow{\mathbf{r}_{0:N}}, \overleftarrow{\mathbf{r}_{N:0}}] \label{eq:decoder_init}
    \\
    \mathbf{a}_{t+1} &=& \text{Attention}(\mathbf{s}_t, [\mathbf{r}_0, \dots, \mathbf{r}_N]) \label{eq:decoder_attention}
    \\
    \mathbf{s}_{t+1}, \mathbf{o}_{t+1} &=& \text{RNN}_{\text{step}}(\mathbf{s}_t, \mathbf{a}_t, W_{pos} \mathbf{p}_t) \label{eq:decoder_step}
    \\
    P(p_{t+1} \mid \dots) &=& \text{Softmax}\left( W_\text{output}[\mathbf{o}_{t+1}, \mathbf{a}_{t+1}] \right) \label{eq:decoder_softmax}
\end{eqnarray}
To alleviate the \emph{exploding gradient problem}~\citep{bengio1994learning}, we use \acp{GRU} in both the forward and backward \ac{RNN}s of the encoder, and in the decoder \ac{RNN}.

\subsection{Beam search}\label{sec:beam_search}
As stated in~\S\ref{sec:problem}, our main goal is to predict $K$ most probable click sequences for a given query and search results.
These $K$ sequences are then used to reason about actual user click behavior,
i.e., sequences of clicks that a user could actually perform for the given query and search results (we call them \textit{probably correct} sequences, see~\S\ref{sec:problem}).

\ac{CSM} defines a probability distribution over infinitely many click sequences.
Extracting $K$ most probable sequences in this case is not straightforward
(since we cannot simply go over all sequences and pick $K$ best ones).
We need a means of generating $K$ most probable sequences without having to calculate the probability for every possible click sequence.
To do that, we suggest to use beam search~\cite{graves2012sequence}.

In our experiments, we use $K \le 1024$ and \emph{beam size} $= K$.
Setting the beam size to $K$ guarantees that the $K$ sequences generated by beam search have the highest probabilities according to \ac{CSM},
i.e., they are indeed most probable click sequences according to \ac{CSM}.
Using a smaller beam size allows us to generate $K$ sequences faster, but does not guarantee that these sequences are the most probable ones.

\subsection{Training}\label{sec:training}
We learn the parameters~$\Theta$ of the \ac{CSM} network (both the encoder and decoder parts) by maximizing the log-likelihood of observed click sequences.
We optimize these parameters using \ac{SGD}.
The learning rates for each parameter are adjusted according to the Adam~\citep{kingma2015adam} algorithm using the default values of $\beta_1 = 0.9$, $\beta_2 = 0.999$ and $\varepsilon = 10 ^ {-8}$.
We also use \emph{gradient clipping}~\citep{pascanu2013difficulty} with the norm set to \num{1} to alleviate the \emph{exploding gradient problem}~\citep{bengio1994learning}, which, as mentioned earlier, GRUs also try to mitigate.


\section{Experimental setup}\label{sec:experimental setup}
In this section we describe our experimental setup.
We start by describing the data we use to conduct our experiments (\S\ref{sec:data}).
Then we discuss our evaluation methodology (\S\ref{sec:evaluation_methodology}),
formulate research questions (\S\ref{sec:research_questions}) and list the experiments we run to answer these research questions (\S\ref{sec:experiments}).

\subsection{Data}\label{sec:data}
We use Yandex Relevance Prediction dataset\footnote{\url{https://academy.yandex.ru/events/data_analysis/relpred2011/} (last visited \today)} released in 2011 by Yandex, the major search engine in Russia.
The dataset consists of \num{146278823} query sessions ordered by time.
We use the first half of the dataset for training \ac{CSM}, and \num{100000} randomly selected query sessions from the second half of the dataset for evaluation.
\citet{borisov2016neural} also use the first half of the dataset for training, which allows a direct comparison with their work.

The statistics about the number of query sessions in the test set split by the number and order of clicks is given in Table~\ref{table:dataset_statistic}.
\begin{table}
    \centering 
    \caption{The number of query sessions in the test set split by the number and order of clicks.
    By ordered click sequences we mean those where a user clicks on results in the order they appear on a SERP.
    The total number of click sequences in the test set is \num{100000}.}
    \label{table:dataset_statistic}

    \begin{tabular}{l r r}
        \toprule
        Number of clicks & Ordered sequences & Unordered sequences \\
        \midrule
        0 & \num{30466} & 0 \\
        1 & \num{46550} & 0 \\
        2 & \num{8851} & \num{2143} \\
        3 & \num{3437} & \num{1856} \\
        4 & \num1564{} & \num{1290} \\
        5 & \num{751} & \num{814} \\
        6 & \num{407} & \num{512} \\
        7 & \num{244} & \num{305} \\
        8 & \num{156} & \num{195} \\
        9 & \num{85} & \num{137} \\
        10+ & \num{73} & \num{164} \\
        \bottomrule
    \end{tabular}
\end{table}

\subsection{Evaluation methodology}\label{sec:evaluation_methodology}
To properly evaluate the proposed \ac{CSM} model, we would need to know a set of all (or at least a sample of) probably correct click sequences
for each test query and search results.
Then we could measure how well the $K$ most probable sequences predicted by \ac{CSM}
describe the properties of the known probably correct sequences.

In practice, however, we observe only one (or a few) of all probably correct click sequences
and, therefore, we cannot even argue about their properties.
The best we can do is to check whether the observed click sequence appears in the list of $K$ sequences predicted by \ac{CSM}.
In particular, we measure recall@K, i.e., the fraction of query sessions for which \ac{CSM} includes the observed click sequence in the list of $K$ most probable sequences.

Since \ac{CSM} is the first model for predicting click sequences, there are no baselines to compare against.
However, we can use as a reference level the percentage of query sessions in which users interact with results in the order these results are presented on a \ac{SERP}.
In our test data, this percentage equals \num{92.73}\%.
This is an upper bound under the assumption that a user scans search results sequentially from top to bottom.
This means that a model, that predicts only click sequences ordered by position, will contain the observed click sequence in the list of $K$ most probable sequences for $\le$ \num{92.73}\% of query sessions.

\smallskip\noindent
\textbf{Task 1.}
Predicting whether a user will click on $\le L$ results is a new task
and, hence, there are no standard metrics to evaluate performance on this task and no existing baselines to compare to.

We propose to evaluate the performance on this task using perplexity and, because this is a classification problem for a fixed L, \ac{AUC}.
Perplexity measures how ``surprised'' a model is upon observing $\le L$ clicks.
\ac{AUC} measures the model's discriminative power.

We use a naive baseline which predicts that a user will make $\le L$ clicks with a constant probability calculated on the training set.
\ac{AUC} of such method is $0.5$.

\smallskip\noindent
\textbf{Task 2.}
Predicting whether a click sequence will be ordered by position is also a new task
and, similarly to Task 1, there are no standard metrics to evaluate the performance on this task and no existing baselines to compare to.

Similarly to Task 1, we use perplexity and \ac{AUC}.
Our naive baseline predicts that a click sequence will be ordered by position with a constant probability calculated on the training set.
\ac{AUC} of such method is also $0.5$, as in Task 1.

\smallskip\noindent
\textbf{Task 3.} Following~\citep{dupret2008user,grotov2015comparative,guo2009click,borisov2016neural,wang2015incorporating}, we evaluate the performance on the standard click prediction task using perplexity,
which measures how ``surprised'' a model is upon observing an unordered set of clicks on search engine results.

We use the following click models as our baselines: \ac{DBN}~\citep{chapelle2009dynamic}, \ac{DCM}~\citep{guo2009efficient}, \ac{CCM}~\citep{guo2009efficient}, \ac{UBM}~\citep{dupret2008user} and \ac{NCM}~\citep{borisov2016neural}.
\citet{borisov2016neural} use the same data for training these click models, which allows us to compare with the result reported in their work.

\subsection{Research questions}\label{sec:research_questions}
We aim to answer the following research questions:

\smallskip\noindent
\begin{enumerate}[itemsep=0pt,topsep=0pt,itemindent=0pt, label=\bfseries RQ\arabic*]
    \item How well does \ac{CSM}, described in \S\ref{sec:method}, predict probably correct click sequences?
        \begin{enumerate}[itemsep=0pt,topsep=0pt]
            \item For how many query sessions, the observed click sequence occurs in the list of $K$ most probable click sequences predicted by \ac{CSM}?
            How fast does this number increase with~$K$?
            \item How well does \ac{CSM} perform for query sessions in which clicks
                \begin{inparaenum}[(i)]
                    \item follow the order in which results are presented on a \ac{SERP}, and
                    \item do not follow the order in which results are presented on a \ac{SERP}.
                \end{inparaenum}
            \item How well does \ac{CSM} perform for query sessions with different number of clicks?
            \item Do $K$ most probable click sequences provide good means to reason about the probability distribution over click sequences predicted by \ac{CSM}?
       \end{enumerate}
    \item How well does \ac{CSM} predict the number of clicks on search results (see Task~1 in \S\ref{sec:prediction tasks})?
    \item How well does \ac{CSM} predict whether or not a user will click on results in the order they are presented on a SERP (see Task~2 in \S\ref{sec:prediction tasks})?
    \item How well does \ac{CSM} predict clicks on search results (see Task~3 in \S\ref{sec:prediction tasks})?
        Does it reach the performance of the state-of-the-art click models?
\end{enumerate}

\subsection{Experiments}\label{sec:experiments}
We design our experiments to answer our research questions.

\smallskip\noindent%
\textbf{E1(a).}
To answer RQ1(a), we measure the percentage of query sessions for which the observed click sequence occurs in the list of $K$ most probable click sequences according to \ac{CSM}.
We use $K = \{1, 2, 3, \dots, 1024\}$.

\smallskip\noindent%
\textbf{E1(b).}
To answer RQ1(b), we measure the percentage of query sessions for which the observed click sequence occurs in the list of $K$ most probable click sequences according to \ac{CSM} separately
\begin{inparaenum}[(i)]
    \item for query sessions in which clicks are ordered by position, and
    \item for query sessions in which clicks are not ordered by position.
\end{inparaenum}
We use $K = \{1, 2, 3, \dots, 1024\}$.

\smallskip\noindent%
\textbf{E1(c).}
To answer RQ1(c), we measure the percentage of query sessions for which the observed click sequence occurs in the list of $K$ most probable click sequences according to \ac{CSM} for query sessions with $\le L$ clicks.
We use $K = \{1, 2, 3, \dots, 1024\}$ and $L=\{1, 2, 3, 4, 5\}$.

\smallskip\noindent%
\textbf{E1(d).}
To answer RQ1(d), we compute the total probability of $K$ most probable click sequences according to \ac{CSM}.
If this probability is close to \num{1}, we conclude that using $K$ most probable click sequences is enough to form a representative empirical distribution over click sequences.
And, thus, $K$ most probable click sequences provide good means to reason about the properties of the probability distribution over click sequences predicted by \ac{CSM}.
If the total probability mass of $K$ most probable click sequences is small, we conclude that using these sequences is not enough to reason about the probability distribution over click sequences predicted by \ac{CSM}.
We use $K = \{1, 2, 3, \dots, 1024\}$.

\smallskip\noindent%
\textbf{E2.}
To answer RQ2, we compute probabilities of clicking on $\le L$ results by marginalizing over the $K$ most probable click sequences according to \ac{CSM} (see Eq.~\ref{eq:estimating probability of <= L clicks}).
We use these probabilities to compute perplexity and \ac{AUC}.
We use $K=1024$ and $L=\{1, 2, 3, 4, 5\}$.

\smallskip\noindent%
\textbf{E3.}
To answer RQ3, we compute the probability that a user will click on results in the order these results are presented on a \ac{SERP} by marginalizing over the $K$ most probable click sequences according to \ac{CSM} (see Eq.~\ref{eq:estimating probability that click sequence will be ordered by position}).
We use this probability to compute perplexity and \ac{AUC}, $K = 1024$.

\smallskip\noindent%
\textbf{E4.}
To answer RQ4, we compute probabilities of clicking on each result by marginalizing over the $K$ most probable click sequences according to \ac{CSM} (see Eq.~\ref{eq:estimating unconditional click probabilities}).
We use these probabilities to compute perplexities for each position and average these perplexity values over positions to obtain the final score.
We use $K = 1024$.

\smallskip\noindent%
In our experiments, we use embeddings of size \num{256}, and the same number of \acp{GRU} in all \acp{RNN}.
We train \ac{CSM} using \acl{SGD} with mini-batches of 64 query sessions and the parameters specified in~\S\ref{sec:training}.


\section{Results}\label{sec:results}
In this section we present the results of the experiments described in \S\ref{sec:experiments} and provide answers to the research questions stated in \S\ref{sec:research_questions}.

\subsection{Experiment 1(a)}\label{sec:results of experiment 1 (a)}
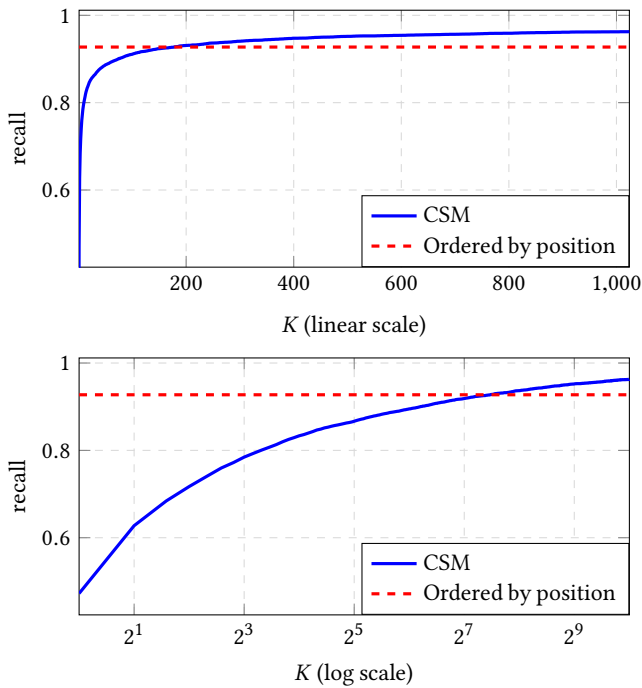
\begin{figure}[h!]
\begin{subfigure}[b]{0.5\textwidth}
    \begin{tikzpicture}
      \begin{axis}[
          width=\linewidth,
          height=50mm,
          grid=major,
          grid style={dashed,gray!30},
          xmin=1,
          xmax=1024,
          xlabel=$K$ (linear scale),
          ylabel=recall,
          ylabel near ticks,
          legend style={at={(1, 0)}, anchor=south east},
          legend cell align=left,
        ]
        \addplot[mark=none, color=blue,very thick] table[x=rank,y=recall,col sep=comma] {data/recalls.csv}; 
        \addplot[color=red, dashed, very thick] coordinates {(1, 0.9273) (1024, 0.9273)};
        \legend{CSM, Ordered by position}
      \end{axis}
    \end{tikzpicture}
    \label{fig:recall at K (linear scale)}
\end{subfigure}
\begin{subfigure}[b]{0.5\textwidth}
    \begin{tikzpicture}
      \begin{axis}[
          width=\linewidth,
          height=50mm,
          grid=major,
          grid style={dashed,gray!30},
          xmode=log,
          log basis x=2,
          xmin=1,
          xmax=1024,
          xlabel=$K$ (log scale),
          ylabel=recall,
          ylabel near ticks,
          legend style={at={(1, 0)}, anchor=south east},
          legend cell align=left,
        ]
        \addplot[mark=none, color=blue,very thick] table[x=rank,y=recall,col sep=comma] {data/recalls.csv}; 
        \addplot[color=red, dashed, very thick] coordinates {(1, 0.9273) (1024, 0.9273)};
        \legend{CSM, Ordered by position}
      \end{axis}
    \end{tikzpicture}
    \label{fig:recall at K (log scale)}
\end{subfigure}
\caption{Percentage of query sessions for which the observed click sequence occurs in the list of $K$ most probable click sequences predicted by \ac{CSM} (blue, solid) and percentage of click sequences ordered by position (red, dashed).}
\label{fig:recall at K}
\end{figure}

Figure~\ref{fig:recall at K} shows recall at different values of $K$ (i.e., the percentage of query sessions for which the observed click sequence occurs in the list of $K$ most probable sequences predicted by \ac{CSM}) in linear and logarithmic scales.
The percentage of query sessions in which clicks are ordered by position equals $92.73$\%.
We show it on the plots as a reference level.

We find that for \num{47.24}\% of query sessions, \ac{CSM} assigns the highest probability to the observed click sequence.
For \num{62.76}\% of query sessions, the observed sequence appears in the list of two sequences with the highest probabilities according to \ac{CSM}.
Since the curve on the logarithmic scale is concave (see Figure~\ref{fig:recall at K}, bottom plot), we conclude that recall of \ac{CSM} increases slower than logarithmically with~$K$.
The percentage of query sessions in which clicks are ordered by position can be seen as an upper bound under the assumption that a user scans search results sequentially from top to bottom.
\ac{CSM} does not make this assumption, and, as a result, is able to reach and surpass this upper bound, achieving \num{96.26}\% recall at $K=1024$.

Answering RQ1(a), we conclude that recall of \ac{CSM} increases slower than logarithmically with $K$, starting from \num{47.24}\% at $K=1$ and reaching \num{96.26}\% at $K=1024$,
which is higher than recall under the sequential assumption ($92.73$\%).

\subsection{Experiment 1(b)}\label{sec:results of experiment 1 (b)}
\begin{figure}[h!]
\begin{subfigure}[b]{0.5\textwidth}
    \begin{tikzpicture}
      \begin{axis}[
          width=\linewidth,
          height=50mm,
          grid=major,
          grid style={dashed,gray!30},
          xmin=1,
          xmax=1024,
          xlabel=$K$ (linear scale),
          ylabel=recall,
          ylabel near ticks,
          legend style={at={(0.5,-.35)},anchor=north},
          legend cell align=left,
        ]
        \addplot[mark=none, color=black,very thick] table[x=rank,y=recall,col sep=comma] {data/recalls.csv}; 
        \addplot[color=black, dashed, very thick] coordinates {(1, 0.9273) (1024, 0.9273)};
        \addplot[mark=none, color=blue,very thick] table[x=rank,y=recall,col sep=comma] {data/recalls_in_ordered_group.csv};
        \addplot[color=blue, dashed, very thick] coordinates {(1, 1.0) (1024, 1.0)};
        \addplot[mark=none, color=red,very thick] table[x=rank,y=recall,col sep=comma] {data/recalls_in_unordered_group.csv};
        \addplot[color=red, dashed, very thick] coordinates {(1, 0.0) (1024, 0.0)};
      \end{axis}
    \end{tikzpicture}
\end{subfigure}

\begin{subfigure}[b]{0.5\textwidth}
    \begin{tikzpicture}
      \begin{axis}[
          width=\linewidth,
          height=50mm,
          grid=major,
          grid style={dashed,gray!30},
          xmode=log,
          log basis x=2,
          xmin=1,
          xmax=1024,
          xlabel=$K$ (log scale),
          ylabel=recall,
          ylabel near ticks,
          legend style={at={(1, -.75)}, anchor=south east},
          legend cell align=left,
        ]
        \addplot[mark=none, color=black,very thick] table[x=rank,y=recall,col sep=comma] {data/recalls.csv}; 
        \addplot[color=black, dashed, very thick] coordinates {(1, 0.9273) (1024, 0.9273)};
        \addplot[mark=none, color=blue,very thick] table[x=rank,y=recall,col sep=comma] {data/recalls_in_ordered_group.csv};
        \addplot[color=blue, dashed, very thick] coordinates {(1, 1.0) (1024, 1.0)};
        \addplot[mark=none, color=red,very thick] table[x=rank,y=recall,col sep=comma] {data/recalls_in_unordered_group.csv};
        \addplot[color=red, dashed, very thick] coordinates {(1, 0.0) (1024, 0.0)};
        \legend{All query sessions, , Query sessions in which clicks ordered by position, , Query sessions in which click not ordered by position}
      \end{axis}
    \end{tikzpicture}
\end{subfigure}

\medskip
\caption{Recall at different values of $K$ for
    (i) all query sessions (black, solid)
    (ii) query sessions in which clicks are ordered by position (blue, solid), and
    (iii) query sessions in which clicks are not ordered by position (red, solid).
Dashed lines show percentages of query sessions in the corresponding groups, in which clicks on results happen in the order these results are presented on a \ac{SERP}.}
    \label{fig:recall at K for ordered and unordered groups}
\end{figure}
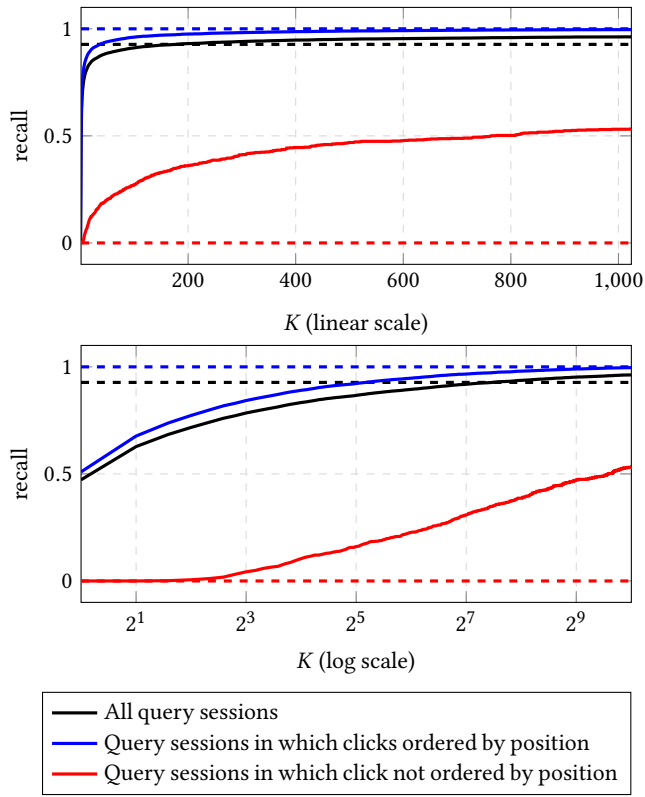

Figure~\ref{fig:recall at K for ordered and unordered groups} shows recall at different values of $K$ (in linear and logarithmic scales) for
\begin{inparaenum}[(i)]
    \item all query sessions (black, solid),
    \item query sessions in which clicks are ordered by position (blue, solid), and
    \item query sessions in which clicks are not ordered by position (red, solid).
\end{inparaenum}
Dashed lines show percentages of query sessions in the corresponding groups, in which clicks on results happen in the order these results are presented on a \ac{SERP}.
Obviously, the second group has 100\% of query sessions with clicks ordered by position (and, hence, the blue dotted line denotes recall of $1$),
while the third group has no such sessions (and, hence, the red dotted line denotes recall of~$0$).

As we find in \S\ref{sec:results of experiment 1 (a)}, \ac{CSM} assigns the highest probability to the observed click sequence for \num{47.24}\% of query sessions.
If we consider only query sessions in which clicks are ordered by position, this percentage goes up to \num{50.94}\%
and then increases slower than logarithmically with $K$ (see Figure~\ref{fig:recall at K for ordered and unordered groups}, bottom plot) achieving \num{99.62}\% at $K = 1024$.
However, if we consider only query sessions in which clicks are not ordered by position, this percentage goes down to \num{0}
and then increases logarithmically with $K$ for $K \ge 5$ (see Figure~\ref{fig:recall at K for ordered and unordered groups}, bottom plot), achieving \num{53.37}\% at $K = 1024$.

It is to be expected that predicting click sequences, where clicks are not ordered by position,
is a much more difficult task than predicting ordered clicks.
First, in our training data the number of ordered click sequences is greater than the number of unordered click sequences
(see Table~\ref{table:dataset_statistic} for the statistic computed on the test set; the training set shares the same distribution).
Second, the number of possible ordered click sequences is less than the number of possible unordered click sequences.
For click sequences of length $L$, there are $N\choose L$ $=$ $\frac{N!}{L!(N - L)!}$ possible ordered click sequences and $N ^ L$ possible unordered click sequences.

And this is where \ac{CSM} makes a difference: in more than 50\% of cases,
the observed click sequence appears in the top $K = 1024$ sequences predicted by \ac{CSM}.
Note that under the assumption that a user scans search results sequentially from top to bottom
such sequences cannot be predicted at all (see the red dotted line at zero recall).
Even in a simple case of predicting ordered click sequences,
\ac{CSM} almost reaches the perfect recall of $1$ for $K = 1024$.

Answering RQ1(b), we conclude that recall of \ac{CSM} is much higher in query sessions in which clicks follow the presentation order than in those in which users click on higher ranked results after clicking on a lower ranked result.

\subsection{Experiment 1(c)}\label{sec:results of experiment 1 (c)}
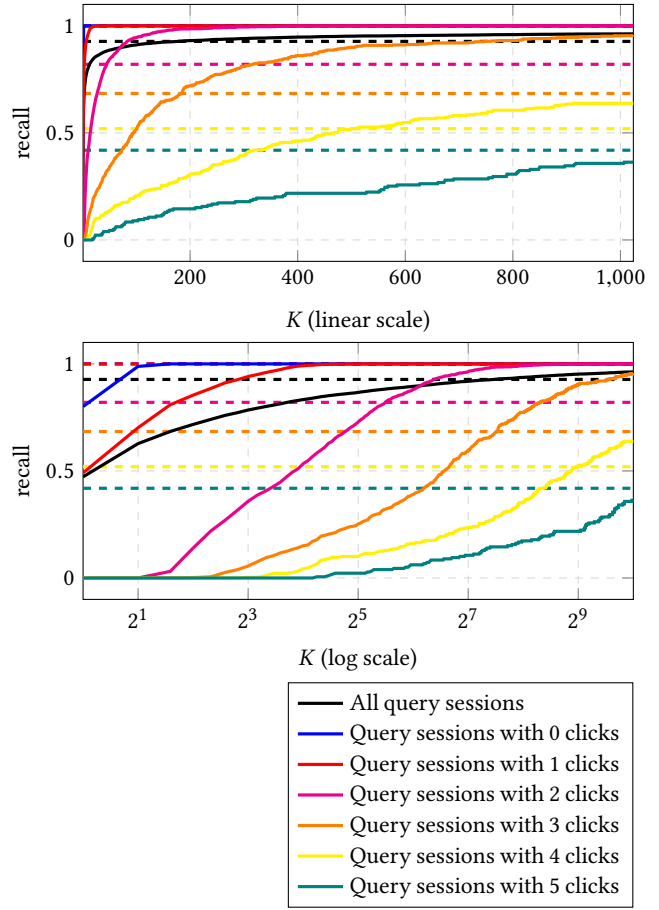
\begin{figure}[h!]
\begin{subfigure}[b]{0.5\textwidth}
    \begin{tikzpicture}
      \begin{axis}[
          width=\linewidth,
          height=50mm,
          grid=major,
          grid style={dashed,gray!30},
          xmin=1,
          xmax=1024,
          xlabel=$K$ (linear scale),
          ylabel=recall,
          ylabel near ticks,
          legend style={at={(1, -1.2)}, anchor=south east},
          legend cell align=left,
        ]
        \addplot[color=black, dashed, very thick] coordinates {(1, 0.9273) (1024, 0.9273)};
        \addplot[color=blue, dashed, very thick] coordinates {(1, 1) (1024, 1)};
        \addplot[color=red, dashed, very thick] coordinates {(1, 1) (1024, 1)};
        \addplot[color=magenta, dashed, very thick] coordinates {(1, 0.8203) (1024, 0.8203)};
        \addplot[color=orange, dashed, very thick] coordinates {(1, 0.6840) (1024, 0.6840)};
        \addplot[color=yellow, dashed, very thick] coordinates {(1, 0.5197) (1024, 0.5197)};
        \addplot[color=teal, dashed, very thick] coordinates {(1, 0.4190) (1024, 0.4190)};
        \addplot[mark=none, color=black,very thick] table[x=rank,y=recall,col sep=comma] {data/recalls.csv}; 
        \addplot[mark=none, color=blue,very thick] table[x=rank,y=recall,col sep=comma] {data/recalls_in_group_0.csv};
        \addplot[mark=none, color=red,very thick] table[x=rank,y=recall,col sep=comma] {data/recalls_in_group_1.csv};
        \addplot[mark=none, color=magenta,very thick] table[x=rank,y=recall,col sep=comma] {data/recalls_in_group_2.csv};
        \addplot[mark=none, color=orange,very thick] table[x=rank,y=recall,col sep=comma] {data/recalls_in_group_3.csv};
        \addplot[mark=none, color=yellow,very thick] table[x=rank,y=recall,col sep=comma] {data/recalls_in_group_4.csv};
        \addplot[mark=none, color=teal,very thick] table[x=rank,y=recall,col sep=comma] {data/recalls_in_group_5.csv};
      \end{axis}
    \end{tikzpicture}
\end{subfigure}

\begin{subfigure}[b]{0.5\textwidth}
    \begin{tikzpicture}
      \begin{axis}[
          width=\linewidth,
          height=50mm,
          grid=major,
          grid style={dashed,gray!30},
          xmode=log,
          log basis x=2,
          xmin=1,
          xmax=1024,
          xlabel=$K$ (log scale),
          ylabel=recall,
          ylabel near ticks,
          legend style={at={(1, -1.2)}, anchor=south east},
          legend cell align=left,
        ]
        \addplot[color=black, dashed, very thick] coordinates {(1, 0.9273) (1024, 0.9273)};
        \addplot[color=blue, dashed, very thick] coordinates {(1, 1) (1024, 1)};
        \addplot[color=red, dashed, very thick] coordinates {(1, 1) (1024, 1)};
        \addplot[color=magenta, dashed, very thick] coordinates {(1, 0.8203) (1024, 0.8203)};
        \addplot[color=orange, dashed, very thick] coordinates {(1, 0.6840) (1024, 0.6840)};
        \addplot[color=yellow, dashed, very thick] coordinates {(1, 0.5197) (1024, 0.5197)};
        \addplot[color=teal, dashed, very thick] coordinates {(1, 0.4190) (1024, 0.4190)};
        \addplot[mark=none, color=black,very thick] table[x=rank,y=recall,col sep=comma] {data/recalls.csv}; 
        \addplot[mark=none, color=blue,very thick] table[x=rank,y=recall,col sep=comma] {data/recalls_in_group_0.csv};
        \addplot[mark=none, color=red,very thick] table[x=rank,y=recall,col sep=comma] {data/recalls_in_group_1.csv};
        \addplot[mark=none, color=magenta,very thick] table[x=rank,y=recall,col sep=comma] {data/recalls_in_group_2.csv};
        \addplot[mark=none, color=orange,very thick] table[x=rank,y=recall,col sep=comma] {data/recalls_in_group_3.csv};
        \addplot[mark=none, color=yellow,very thick] table[x=rank,y=recall,col sep=comma] {data/recalls_in_group_4.csv};
        \addplot[mark=none, color=teal,very thick] table[x=rank,y=recall,col sep=comma] {data/recalls_in_group_5.csv};
        \legend{,,,,,,,All query sessions, Query sessions with $0$ clicks, Query sessions with $1$ clicks, Query sessions with $2$ clicks, Query sessions with $3$ clicks, Query sessions with $4$ clicks, Query sessions with $5$ clicks}
      \end{axis}
    \end{tikzpicture}
\end{subfigure}

\medskip
\caption{Recall at different values of $K$ for (i) all query sessions and (ii) query sessions with $L$ clicks.
    Dashed lines show percentages of query sessions in the corresponding groups, in which clicks on results happen in the order these results are presented on a \ac{SERP}.}
    \label{fig:recall at K for query sessions with L clicks}
\end{figure}

Figure~\ref{fig:recall at K for query sessions with L clicks} shows recall at different values of $K$ for
\begin{inparaenum}[(i)]
\item all query sessions and
\item query sessions with $L$ clicks.
\end{inparaenum}
Dashed lines show percentages of query sessions in the corresponding groups, in which clicks on results happen in the order these results are presented on a \ac{SERP}.

For click sequences of length $L = 0$ and $L = 1$, recall of \ac{CSM} approaches \num{1} (already for small values of $K$).
Recall of CSM for sequences of length $L=2, 3, 4$ at $K=1024$ is higher than the percentages of query sessions in which click sequences of length~$L$ are ordered by position.
For $L=5$ and $K=1024$, recall of CSM approaches the percentage of query sessions in which click sequences are of length \num{5} and are ordered by position.
For sequences of length $\ge 2$, recall of \ac{CSM} first increases logarithmically with $K$ (for $K \ge K_0(L)$),
and then might increase both faster and slower then logarithmically with $K$ depending on $L$ and the range of $K$.

We can see that the longer a click sequence is, the more difficult it is to predict such a sequence.
This is intuitive and can be explained similarly to \S\ref{sec:results of experiment 1 (b)}.
First, we have more training data for shorter click sequences (see Table~\ref{table:dataset_statistic}).
Second, the number of possible click sequences of length~$L$ increases exponentially with $L$,
making the prediction task more difficult.

Answering RQ1(c), we conclude that recall of \ac{CSM} is very high in query sessions with a small number of clicks, and lower in query sessions with a larger number of clicks.

\subsection{Experiment 1(d)}\label{sec:results of experiment 1 (d)}
\begin{figure}[h!]
    \begin{tikzpicture}
      \begin{axis}[
          width=\linewidth,
          height=50mm,
          grid=major,
          grid style={dashed,gray!30},
          xmin=1,
          xmax=1024,
          xlabel=$K$,
          ylabel=$\sum_{i=1}^{K} P_\text{CSM}(s_i)$,
          ylabel near ticks
        ]
        \addplot[mark=none, color=blue,very thick] table[x=k,y=total_probability,col sep=comma] {data/total_probability_of_k_most_probable_sequences.csv}; 
      \end{axis}
    \end{tikzpicture}
    \caption{Total probability of $K$ most probable sequences predicted by \ac{CSM} for different values of~$K$.}
    \label{fig:probability_of_k_most_probable_sequences}
\end{figure}
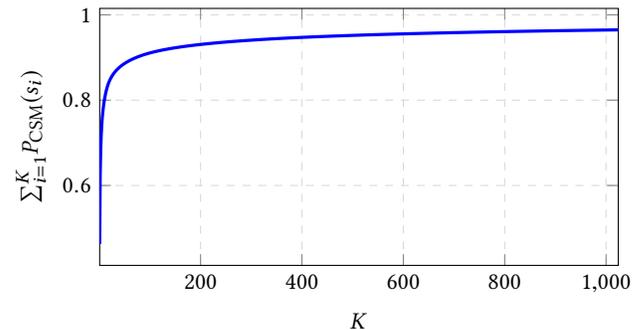

Figure~\ref{fig:probability_of_k_most_probable_sequences} plots, for different values of $K$, the total probability of $K$ most probable click sequences predicted by \ac{CSM} averaged over query sessions in the test set.
We write $\sum_{i=1}^{K} P_\text{CSM}(s_i)$ to denote this probability.

We find that the total probability of $K$ most probable click sequences grows fast with $K$, starting from $46.29$\% at $K=1$ and $71.69$\% at $K=4$, and achieving $91.81$\% at $K=128$ and $96.47$\% at $K=1024$.
Thus, even for modest values of $K$, the total probability of the $K$ most probable click sequences is not significantly less than~\num{1}.

Answering RQ1(d), we conclude that the $K$ most probable click sequences according to \ac{CSM} provide good means to reason about the whole probability distribution over click sequences predicted by \ac{CSM}.

\subsection{Experiment 2}\label{sec:results of experiment 3}
The results on Task 1 (\S\ref{sec:prediction tasks}) are given in Tables~\ref{table:perplexities of <= L clicks} and~\ref{table:auc for the task of predicting whether a user will click <= L results}.
Table~\ref{table:perplexities of <= L clicks} shows perplexity of \ac{CSM} upon observing a sequence of $\le L$ clicks.
Table~\ref{table:auc for the task of predicting whether a user will click <= L results} shows \ac{AUC} of \ac{CSM} on the same task.
Recall, that the naive baseline predicts that a user will make $\le L$ clicks with a constant probability optimized on the training set.

\begin{table}[h]
    \caption{Perplexity of observing a sequence of $\le L$ clicks.
     Lower values correspond to better prediction performance.}
    \centering
    \begin{tabular}{@{} l r r r r r r r @{}}
        \toprule
        & \multicolumn{1}{c}{$L=0$} &\multicolumn{1}{c}{$L \le 1$} & \multicolumn{1}{c}{$L \le 2$} & \multicolumn{1}{c}{$L \le 3$} & \multicolumn{1}{c}{$L \le 4$} & \multicolumn{1}{c}{$L \le 5$} \\
        \midrule
        Baseline & $1.8512$ & $1.7169$ & $1.4450$ & $1.2779$ & $1.1784$ & $1.1068$ \\
        \ac{CSM} & $1.7155$ & $1.6153$ & $1.3852$ & $1.2438$ & $1.1602$ & $1.1029$ \\
        \bottomrule
    \end{tabular}
    \label{table:perplexities of <= L clicks}
\end{table}

\begin{table}[h]
    \caption{\ac{AUC} for the task of predicting whether a user will click on $\le L$ results.}
    \centering
    \begin{tabular}{@{} l r r r r r r r @{}}
        \toprule
        & \multicolumn{1}{c}{$L=0$} &\multicolumn{1}{c}{$L \le 1$} & \multicolumn{1}{c}{$L \le 2$} & \multicolumn{1}{c}{$L \le 3$} & \multicolumn{1}{c}{$L \le 4$} & \multicolumn{1}{c}{$L \le 5$} \\
        \midrule
        Baseline & \num{.5000} & \num{.5000} & \num{.5000} & \num{.5000} & \num{.5000} & \num{.5000} \\
        \ac{CSM} & \num{.7362} & \num{.7278} & \num{.7353} & \num{.7535} & \num{.7566} & \num{.7795} \\
        \bottomrule
    \end{tabular}
    \label{table:auc for the task of predicting whether a user will click <= L results}
\end{table}

Both Tables~\ref{table:perplexities of <= L clicks} and \ref{table:auc for the task of predicting whether a user will click <= L results}
show that \ac{CSM} predicts the number of clicks better than the baseline
and its performance increases with $L$ (lower perplexity and higher \ac{AUC}).
The latter result is intuitive as more sequences have $\le L$ clicks for larger $L$
and, thus, the prediction task becomes easier as $L$ grows.

Answering RQ3, we conclude that \ac{CSM} provides good means to predict the number of clicked results.

\subsection{Experiment 3}\label{sec:results of experiment 4}
The results on Task 2 (\S\ref{sec:prediction tasks}) are given in Table~\ref{table:perplexity of observing clicks ordered by position}.
The table shows perplexity and \ac{AUC} of \ac{CSM} when predicting a sequence of clicks ordered by position.

\begin{table}[h]
    \caption{Performance for the task of predicting whether a user will click on results in the order these results are presented on a \ac{SERP}.
    Lower values of perplexity and larger values of \ac{AUC} correspond to better prediction performance.}
    \centering
    \begin{tabular}{ l r r }
        \toprule
        & Perplexity & \ac{AUC} \\
        \midrule
        Baseline & \num{1.2984} & \num{.5000} \\
        \ac{CSM} & \num{1.2788} & \num{.6826} \\
        \bottomrule
    \end{tabular}
    \label{table:perplexity of observing clicks ordered by position}
\end{table}

Table~\ref{table:perplexity of observing clicks ordered by position} shows that \ac{CSM} outperforms the baseline in terms of both perplexity and \ac{AUC}.
Thus, answering RQ3, we conclude that \ac{CSM} provides good means to predict whether a user will interact with results in the order these results are presented on a SERP.

\subsection{Experiment 4}\label{sec:results of experiment 2}
The results on Task 3 (\S\ref{sec:prediction tasks}), i.e., the click prediction task, are given in Table~\ref{table:perpexity on the click prediction task}.
The results for \ac{DBN}, \ac{DCM}, \ac{CCM}, \ac{UBM} and \ac{NCM} are according to~\citep{borisov2016neural}.
\begin{table}[h]
    \caption{Perplexity for the click prediction task.
    Lower values correspond to better prediction performance.
    The results for \ac{DBN}, \ac{DCM}, \ac{CCM}, \ac{UBM} and \ac{NCM} are according to~\citep{borisov2016neural}.}
    \centering
    \begin{tabular}{ l r }
        \toprule
        Click model & Perplexity \\
        \midrule
        DBN & $1.3510$ \\
        DCM & $1.3627$ \\
        CCM & $1.3692$ \\
        UBM & $1.3431$ \\
        NCM & $1.3318$ \\
        \midrule
        \ac{CSM} & $1.3312$\\
        \bottomrule
    \end{tabular}
    \label{table:perpexity on the click prediction task}
\end{table}

Table~\ref{table:perpexity on the click prediction task} shows that \ac{CSM} outperforms \ac{DBN}, \ac{DCM}, \ac{CCM} and \ac{UBM} by a large margin and matches the performance of \ac{NCM}, which is reported to be the state-of-the-art click model~\citep{borisov2016neural}.
Answering RQ4, we conclude that \ac{CSM} provides good means to predict clicks on search engine results, achieving the state-of-the-art performance.


\section{Related Work}
\label{sec:related work}
We describe two types of related work: user interactions in search and user modeling.

\subsection{User interactions in search}

Log data from interactive systems such as search engines is one of the most ubiquitous forms of data available, as it can be recorded at little cost~\citep{white-interactions-2016}.
These data have become a popular source for improving the performance of search engines.
In particular, logged interactions have been successfully adopted to improve various aspects of search, including document ranking~\citep{joachims-optimizing-2002,agichtein2006improving,o2016leveraging}, query auto-completion~\citep{li2014two,jiang2014learning} and query suggestion~\citep{cao2008context,wu2013learning}, to improve recommender systems~\citep{oard-implicit-1998}, optimizing presentations~\cite{wang2016beyond}, and evaluation~\citep{hofmann-estimating-2012}.

In the context of web search, many types of implicit information interaction behavior have been studied over the years.
Early work, e.g., by \citet{craswell2008experimental}, focuses on single clicks and, in particular, on the \emph{first} click.
And assumes that a user abandons examination of web results upon the first click.
\citet{guo2009efficient} expand on this by studying sessions with \emph{multiple} clicks, looking not just at the first click but also at follow-up clicks, the last click and dependencies between clicks, reflecting more complex information behavior.

There is a very broad spectrum of research that studies and tries to interpret information interaction behavior that involves multiple clicks, either by also taking additional signals into consideration or by zooming in on specific aspects of sequences of clicks.
Examples of the former include work by \citet{huurnink-search-2010} who examine click signals, download behavior, and purchase signals in vertical search and find high degrees of correlation between the three.
Time, such as dwell time or time between user actions such as clicks, has been found to be another important source of implicit signals~\citep{borisov2016context}: times elapsed between user actions provide means to measure user satisfaction at the result level~\citep{fox2005evaluating,kim2014modeling}, session level~\citep{fox2005evaluating,hassan2012semi} and system level~\citep{chapelle2012large,schuth_2015_predicting}.
And beyond that, on mobile or screen-less devices there is range of interaction signals that are different from signals familiar from desktop environment -- due to the context of use and due to gesture- and voice-based control, such as swipes, touch and voice conversations -- and that have not been studied extensively~\citep{kiseleva-evaluating-2017}. 
Our work differs from these publications as we remain focused on click signals only and especially on sequences click signals.

Relevant examples of the studies that zoom in on specific aspects of multiple click behavior include work on repeat behavior such as repeated examinations or clicks~\citep{oard-implicit-1998,xu2012incorporating}, which can be interpreted as strong evidence of the value ascribed to the result being examined or clicked again.
In a similar vein, \citet{scaria-last-2014} consider back clicks and last clicks; in their view, back clicks suggest a lack of progress on the current navigational path and, depending on contextual factors, last clicks mark success or failure. 
\citet{hassan-awadallah-struggling-2014} and \citet{odijk-struggling-2015} focus on aspects of click sequences, including the number of clicks, their dwell time, and features to capture whether the user was clicking on the same results or results from the same domain multiple times, indicative of difficulty locating a particular resource.
\citet{williams-does-2017} examine whether sequences of user interactions over time can be used to differentiate between good and abandonment and train an LSTM to distinguish between the two types of behavior.
Especially relevant for our paper is the work by \citet{wang2015incorporating}, who consider non-sequential examination and click behavior, both through an eye-tracking study and a log-based study.
They arrive at several behavioral principles, for instance%
\begin{inparaenum}[(i)]
    \item between adjacent clicks, users tend to examine search results in a single direction without changes, and the direction is usually consistent with that of clicks; and
    \item although the examination behavior between adjacent clicks can be regarded as locally unidirectional, users may skip a few results and examine a result at some distance from the current one following a certain direction.
\end{inparaenum}

\subsection{Modeling user interactions}
To understand, describe and predict various types of user interactions discussed above,
a number of user interaction models have been proposed aimed at modeling clicks~\cite{chuklin2015click},
mouse movements~\cite{diaz2013robust}, dwell time~\cite{kim2014modeling,liu2010understanding}, etc.

So far, modeling user clicks in search has attracted the most attention~\cite{chuklin2015click}.
Click models usually represent clicks as binary random variables and construct a \ac{PGM}
that describes the dependencies between clicks and other (usually hidden) random variables,
such as attractiveness (i.e., whether a snippet is attractive to a user given a query)
and examination (i.e., whether a snippet is examined by a user)~\cite{chapelle2009dynamic,craswell2008experimental,dupret2008user,guo2009click,guo2009efficient}.
The advantage of \ac{PGM}-based click models is that they intuitively describe user click behavior and can predict future clicks based on past observations~\cite{chuklin2015click}.
Some click models take into account the order in which a user interacts with the results
in order to better model and predict clicks~\cite{xu2010temporal,wang2010inferring,wang2015incorporating,liu2016time,xie-constructing-2018}.
However, such models either do not aim at predicting click sequences~\cite{wang2010inferring,wang2015incorporating,liu2016time,xie-constructing-2018} or consider only very short sequences of clicks~\cite{xu2010temporal}.

Recently, neural click models have been proposed~\cite{zhang2014sequential,borisov2016neural}.
The advantage of these models is that they do not require manually constructed \ac{PGM}s to describe and predict user clicks,
but rely on raw click data to learn hidden click patterns.
Neural click models have better click prediction accuracy, but suffer from uninterpretability of the learned neural model as opposed to easily interpretable \ac{PGM}-based click models.
Also, as before, neural click models cannot predict sequences of clicks.

In addition to clicks, mouse movements between search results and various search-related timings have been studied and modeled.
The probability of hovering over one element of a \ac{SERP} after hovering over another element is predicted using the Farley-Ring model in~\cite{diaz2013robust}.
Dwell time is modeled though Weibull and gamma distributions in~\cite{kim2014modeling,liu2010understanding}.
More timings, such as time between clicks, time to first/last click, etc., are considered in~\cite{borisov2016context},
where, in addition to the above distribution-based models, a context bias is modeled using neural networks.

\smallskip\noindent
What our work adds on top of the work listed above is our focus on \emph{sequences} of clicks, and in particular on describing a set of \emph{probably correct} click sequences.


\section{Conclusion and future work}\label{sec:conclusion and future work}
In this paper, we studied the problem of predicting sequences of user interactions and, in particular, sequences of clicks.
We formally defined the problem of click sequence prediction and introduced the notion of probably correct click sequences.
Furthermore, we proposed \ac{CSM}, a neural network based model for predicting a probability distribution over click sequences.
We advocated for using the $K$ most probable click sequences predicted by \ac{CSM} as a set probably correct click sequences.
And suggested to use these $K$ click sequences to reason about the properties of the probability distribution over click sequences, such as the expected number of clicks and the expected order of clicks.

We evaluated the quality of \ac{CSM} on a publicly available dataset.
First, we showed that even for modest thresholds the $K$ (larger than the threshold) most probable click sequences predicted by the \ac{CSM} constitute a substantial part of the total probability mass assigned by the \ac{CSM} to all possible click sequences, and thus can be regarded as probably correct click sequences predicted by \ac{CSM}.
We proposed to judge the success of a click sequence model~$\mathcal{M}$ by the fact that the observed click sequence occurs in the list of the $K$ most probable click sequences predicted by $\mathcal{M}$.
We measured performance of \ac{CSM} using recall@K, a metric that is also used to evaluate the performance of approximate nearest neighbor search methods~\citep{muja2009fast,jegou2011product}.
Our results showed that recall@K grows fast with $K$, starting from \num{47.24}\% at $K = 1$ and reaching \num{96.26}\% at $K = 1024$.
We also found that recall@K increases slower with $K$ in query sessions with larger number of clicks and in query sessions where users click on higher ranked results after clicking on a lower ranked result.

We also evaluated \ac{CSM} on three prediction tasks:
\begin{inparaenum}[(i)]
    \item predicting the number of clicks,
    \item predicting non-consecutive click sequences, and
    \item predicting clicks.
\end{inparaenum}
The first two tasks were proposed in our work for the first time and the last one is a standard task used to evaluate click models.
We found that \ac{CSM} shows reasonable performance on the first two tasks, outperforming naive baselines that predict
\begin{inparaenum}[(i)]
    \item that a user will click on $\le L$ results, and
    \item that a user will click on the results in a non-consecutive order
\end{inparaenum}
with a constant probability optimized on the training set.
Finally, we observed that \ac{CSM} reaches state-of-the-art performance on the task of predicting clicks, outperforming \ac{PGM}-based click models \ac{DBN}~\citep{chapelle2009dynamic}, \ac{DCM}~\citep{guo2009efficient}, \ac{CCM}~\citep{guo2009efficient} and \ac{UBM}~\citep{dupret2008user} by a large margin, and matching the results of the recently proposed \ac{NCM}~\citep{borisov2016neural}, which is also implemented as a neural network.

In contrast to previous studies, which focus on modeling and predicting separate interaction events (e.g., a click on a result or mouse movement between two results) and properties of these separate events (e.g., time between clicks), our work focuses on understanding, modeling and predicting sequences of these events.

As to future work, we see two main directions:%
\begin{inparaenum}[(i)]
    \item to consider other representations of the user's query and the results returned by a search engine, and
    \item to extend \ac{CSM} to non-linear \ac{SERP} layouts.
\end{inparaenum}
The user's query can be represented by its text, and the results by their content (title, snippet and main content).
We believe that using content-based representations will allow us to learn more interesting dependencies between the results, and improve the performance for rare queries.
The encoder proposed in \S\ref{sec:network_architecture} makes use of the fact that search results are presented as a list.
Recommender systems present their results using non-linear layouts.
Generalizing the encoder will make \ac{CSM} suitable for applications outside of web search.

\subsection*{Acknowledgements}
This research was partially supported by
Ahold Delhaize,
Amsterdam Data Science,
the Bloomberg Research Grant program,
the China Scholarship Council,
the Criteo Faculty Research Award program,
Elsevier,
the European Community's Seventh Framework Programme (FP7/2007-2013) under
grant agreement nr 312827 (VOX-Pol),
the Google Faculty Research Awards program,
the Microsoft Research Ph.D.\ program,
the Netherlands Institute for Sound and Vision,
the Netherlands Organisation for Scientific Research (NWO)
under pro\-ject nrs
CI-14-25, 
652.\-002.\-001, 
612.\-001.\-551, 
652.\-001.\-003, 
and
Yandex.
All content represents the opinion of the authors, which is not necessarily shared or endorsed by their respective employers and/or sponsors.

\bibliographystyle{ACM-Reference-Format}
\balance
\bibliography{sigir2018-fp-nsncm}

\end{document}